\documentclass{PoS}

\title{New Long-Lived Particles at the LHC}

\ShortTitle{New Long-Lived Particles at the LHC}

\author{Daniel Stolarski\\
        Ottawa-Carleton Institute for Physics, Carleton University \\ 
        1125 Colonel By Drive,  Ottawa, Ontario K1S 5B6, Canada\\
        E-mail: \email{stolar@physics.carleton.ca}}

\abstract{I discuss various scenarios for long-lived particles at the LHC that can have spectacular signatures, not all of which have been searched for. I take motivation from dark matter and the hierarchy problem, as well as from finding novel signatures. There is also a brief discussion of triggers.}

\FullConference{The European Physical Society Conference on High Energy Physics\\
		5-12 July, 2017\\
		Venice}

\begin{document}

\section{Keys to Longevity}
In this talk, I describe various scenarios which predict new long-lived particles at the LHC. These come from many types of physics beyond the Standard Model (BSM), but I will begin with a more general discussion of what can cause particles to live a long time. 

Particle longevity is defined in terms of the width $\Gamma$ of a particle in units of its mass, namely $\Gamma/m \ll 1$. There are at lest three ways for a particle to live a long time. The first is from \textbf{heavy scales}. If the decay of a particle of mass $m$ is mediated by a much heavier field of mass $M_*$, then we have
\begin{equation}
\Gamma \sim (M/M_*)^\#
\end{equation}
where $\#$ is some positive integer that depends on the details of the model. An example from the Standard Model (SM) is the muon, which has $\Gamma/m \simeq 3\cdot 10^{-18}$ because the decay is mediated by the $W$ boson which is much heavier than the muon. An example from BSM physics is proton decay in Grand Unified Theories (GUT). These models predict new fields that violate baryon and lepton number, but the mass of those fields is typically $\sim 10^{16}$ GeV, so these models predict a proton lifetime on the scale of $10^{30}$ years.

A second way that a particle can live a long time is via \textbf{small couplings}. An example from the SM is to compare the decay of $b$-mesons to $c$-mesons. Both of course are mediated by the $W$, but the arguments of the previous paragraph would indicate that the charm mesons should live much longer. This is not what is observed because bottom meson decays are proportional to the square of the CKM element $V_{cb} \simeq 0.04$, while charm decays do not have any such suppression: their decays go as $V_{cs}\simeq 0.97$. An example from BSM physics is in $R$-parity violating (RPV) supersymmetry (SUSY). In that case, because the RPV couplings are in the superpotential, it is natural for them to be quite small, which would in turn give the lightest SUSY particle (LSP) a long lifetime.

A third way to generate a long lifetime is with \textbf{kinematic squeezing}: if the sum of the masses of the decay products is very close to the mass of the parent particle, then the decay can be quite slow. An example from the SM is neutron decay, the neutron lives nearly 15 minutes. The neutron decays to a proton, an electron, and a neutrino, and, even ignoring the neutrino mass, we have
\begin{equation}
m_n - m_p -m_e \ll m_n
\end{equation}
so the neutron lifetime is long. An example from BSM models are pure Higgsinos in SUSY models. The charged Higgsinos gets a small loop induced mass splitting from the neutral one, therefore making them nearly degenerate. A fourth, somewhat more exotic way that particles can live a long time is discussed in Sec.~\ref{sec:quirks}.

\section{Motivation}

The LHC experiments have performed many searches for long-lived particles. Some of the results can be found in~\cite{ATLAS:LLP,CMS:LLP}, as well as in talks in this conference by Mauri, Lusiani, Adams, Petterson, Lutz, Saito, and Otono. As I will discuss below, long-lived particles comprise many different possible signatures, not all of which have been searched for. Furthermore, the lack of any evidence for physics beyond the SM at the LHC has put BSM theories under significant strain. Yet, the motivations to look for physics beyond the SM are as strong as ever. Therefore, we must continue to explore the space of models to make sure that if BSM physics is produced at the LHC, that it would be discovered. In the following sections, I will discuss several different types of signatures for long-lived particles, some of which arise from strongly motivated models, and others which are discussed for the uniqueness of the experimental signature.

\section{Dark Matter}

There is extremely convincing evidence for dark matter on many length scales, but all the observations are of the gravitational interactions of dark matter. As of yet, we have no information about the particle physics properties of dark matter, despite many very sophisticated experimental searches. The searches for Weakly Interacting Massive Particles (WIMP), in particular, have made tremendous progress exploring the parameter space, and yet have thus far come up empty. Here, I will consider two dark matter paradigms that differ from WIMPs and give rise to new long-lived particle signatures at the LHC.

\subsection{Freeze-In}

In the freeze-in scenario~\cite{Hall:2009bx}, dark matter is never in thermal equilibrium with the bath of SM particles. There does exist a very rare process where dark matter particles are produced from the thermal bath. Therefore, the SM slowly leaks energy from the bath to the dark matter sector until the dark matter abundance is ``frozen in.'' The abundance is thus set by a small coupling, and, unlike for freeze out, the larger the coupling, the larger the abundance of dark matter. This process can be written as 
\begin{equation}
B \rightarrow A_{SM} X \; ,
\end{equation}
where $X$ is the dark matter, $A_{SM}$ is an SM set of states, and $B$ is a BSM state with large couplings to the SM and small couplings to the dark matter. This implies that if it is kinematically accessible, $B$ can be produced copiously at the LHC. Furthermore, because of its tiny coupling to dark matter, it will naturally be long lived. When it decays, it will produce $A_{SM}$, which could be any SM set of states. 

This scenario was explored in more detail in~\cite{Co:2015pka}. This framework can give rise to all kinds of displaced particle signatures including displaced jets or leptons. Furthermore, $B$ could be charged under electromagnetism or colour, so this also could give rise to stopped charged particles and heavy highly ionizing tracks. Thus, this scenario can motivate the search for any type of SM object appearing in any of the sub detector components of the LHC experiments. 

\subsection{Asymmetric Dark Matter and Emerging Jets}

The energy density of dark matter in the universe is approximately five times the energy of baryons, 
\begin{equation}
\Omega_{DM} \simeq 5 \Omega_{B}\,.
\label{eq:dm}
\end{equation}
The ratio of these two energies could have been orders of magnitude larger or smaller, so the fact that they are comparable is puzzling. Furthermore, this is not explained in WIMP models. 

For baryons, $\Omega_B = m_p n_B$ where $m_p$ is the proton mass which is set by the (known) QCD dynamics. $n_B$ is the number density of baryons which is controlled by the baryon asymmetry, the amount more matter than anti-matter in the universe. Where this asymmetry arises is unknown, but the models that can do this are collectively known as baryogenesis. For dark matter, the same equation holds, $\Omega_{DM} = m_{DM} n_{DM}$. The idea of asymmetric dark matter is that $n_{DM}$ is controlled by the same kind of physics as $n_{B}$. In this case, there is dark matter and anti-dark matter, and more of one than the other. If the same dynamics controls both asymmetries, then it is natural to expect $n_{DM} \simeq n_{B}$. This is an old idea~\cite{Nussinov:1985xr} which has experienced a recent revival~\cite{Kaplan:2009ag} and is reviewed in~\cite{Petraki:2013wwa}. 

In order to completely address the coincidence of the energy density of dark matter and baryons of Eq.~(\ref{eq:dm}), a model must also explain why the mass of the dark matter is similar to the proton mass. This can be done if dark matter is the lightest baryon of a ``dark QCD'' which confines much like the SM QCD~\cite{Bai:2013xga}. In order to produce the requisite asymmetries, these models also require bifundamental scalars charged under both QCD and dark QCD and predict that these scalars are at the TeV scale, as opposed to all the dark hadrons which are at the few GeV scale. 

\begin{figure}
\begin{centering}
\includegraphics[width=0.5\textwidth]{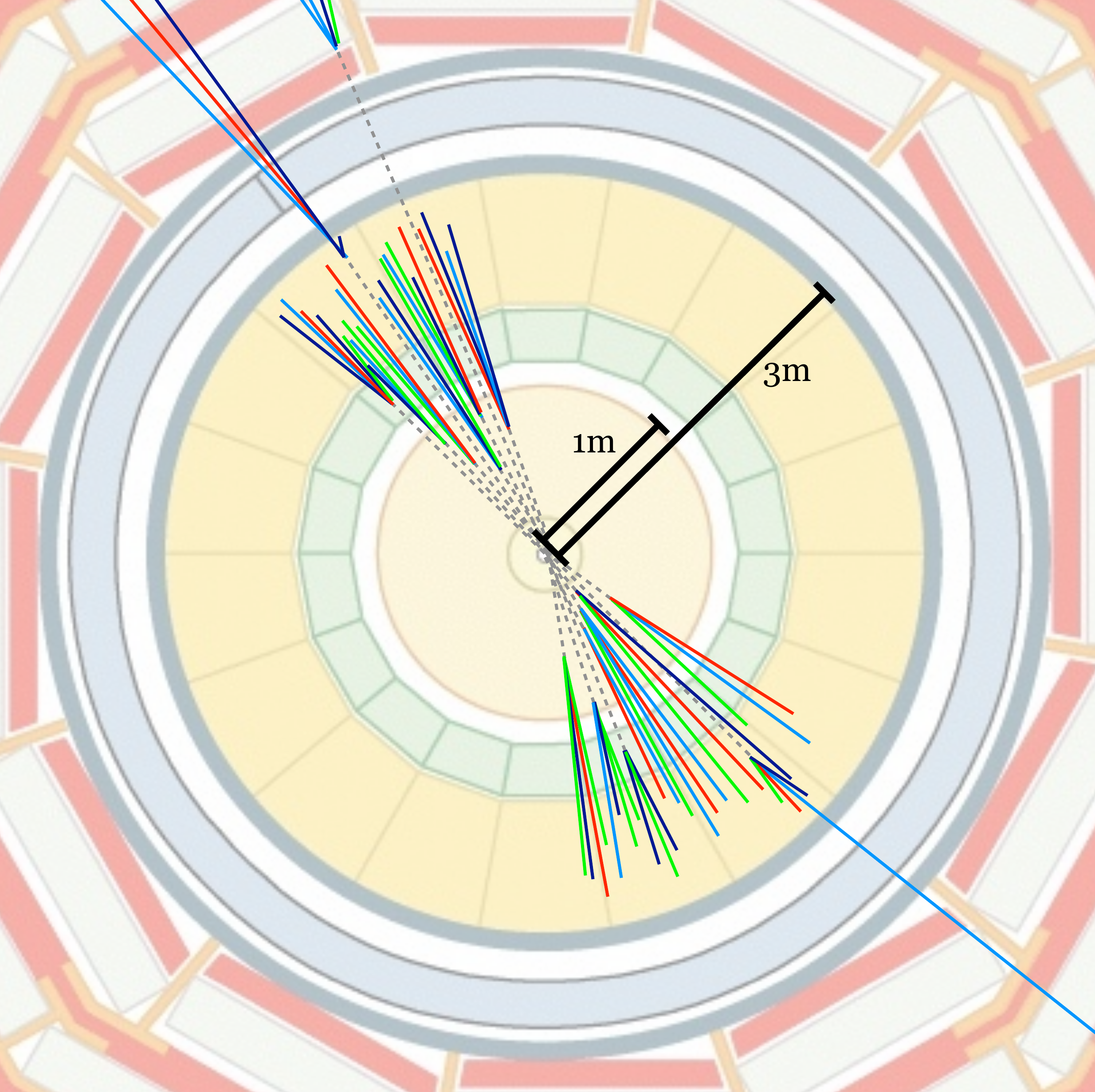}
\caption{Figure 1 from~\cite{Schwaller:2015gea}. A pictorial representation of an event with two emerging jets. }
\label{fig:emerging}
\end{centering}
\end{figure}

This heavy mediator can be produced at the LHC and will decay to a quark and a dark quark. Both quarks will hadronize and form jets in their respective sectors. The dark mediator will also mediate the decay of dark mesons, and because it is much heavier than the dark confinement scale, the dark mesons will have macroscopic lifetimes. So when a dark quark is produced, it will shower and then hadronize to many invisible dark hadrons, most which will slowly decay back to the SM, so that at long distance, this will look like an ordinary jet. This signature is termed ``emerging jets''~\cite{Schwaller:2015gea}, and is shown pictorially in Fig.~\ref{fig:emerging}. The LHC experiments have significant reach to search for these kinds of spectacular events, but no currently published search would be sensitive. 

\newpage

\section{Other Confining Dark Sectors}

The idea of having a dark QCD-like confining sector arises in many contexts besides asymmetric dark matter and can give rise to a variety of different signatures.

\subsection{Twin Higgs and Folded SUSY}

The gauge hierarchy problem has been the driving force in much of the study of BSM physics: how do you tame radiative corrections to the Higgs mass that are quadratically sensitive to high scales? This is typically done with new fields to cancel the radiative corrections from the SM fields. Until relatively recently, the standard lore was that some of the new fields had to be coloured under (SM) QCD, but it turns out this is not necessary. Twin Higgs models~\cite{Chacko:2005pe} have fermionic top partners which are uncoloured, and folded SUSY~\cite{Burdman:2006tz} models have scalar top partners. These models still need the factor of $N_c=3$ in the loop correction, so they typically have a twin colour force which confines at a scale slightly higher than the QCD scale. 

\begin{figure}
\begin{centering}
\includegraphics[width=0.5\textwidth]{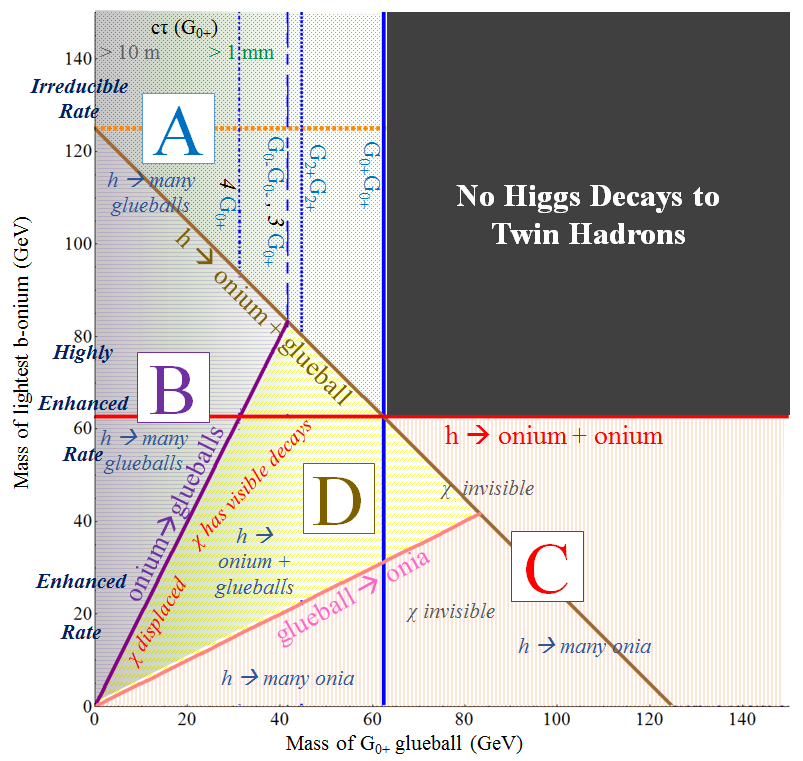}
\caption{Figure 7 from~\cite{Craig:2015pha}. A schematic region of the different parameter regions allowed for exotic Higgs decays in twin Higgs models. }
\label{fig:glueball}
\end{centering}
\end{figure}

The phenomenology of these types of neutral naturalness theories is in some ways similar to emerging jets, but the portal is the SM Higgs. There also tend to be mirror quarks which are somewhat heavier than the QCD scale, the twin-bottom, which can lead to interesting -onium phenomenology. This was studied in detail in~\cite{Craig:2015pha,Curtin:2015fna}, and is shown schematically in Fig.~\ref{fig:glueball}.

\subsection{Quirks}
\label{sec:quirks}

\begin{figure}
\begin{centering}
\includegraphics[width=0.5\textwidth]{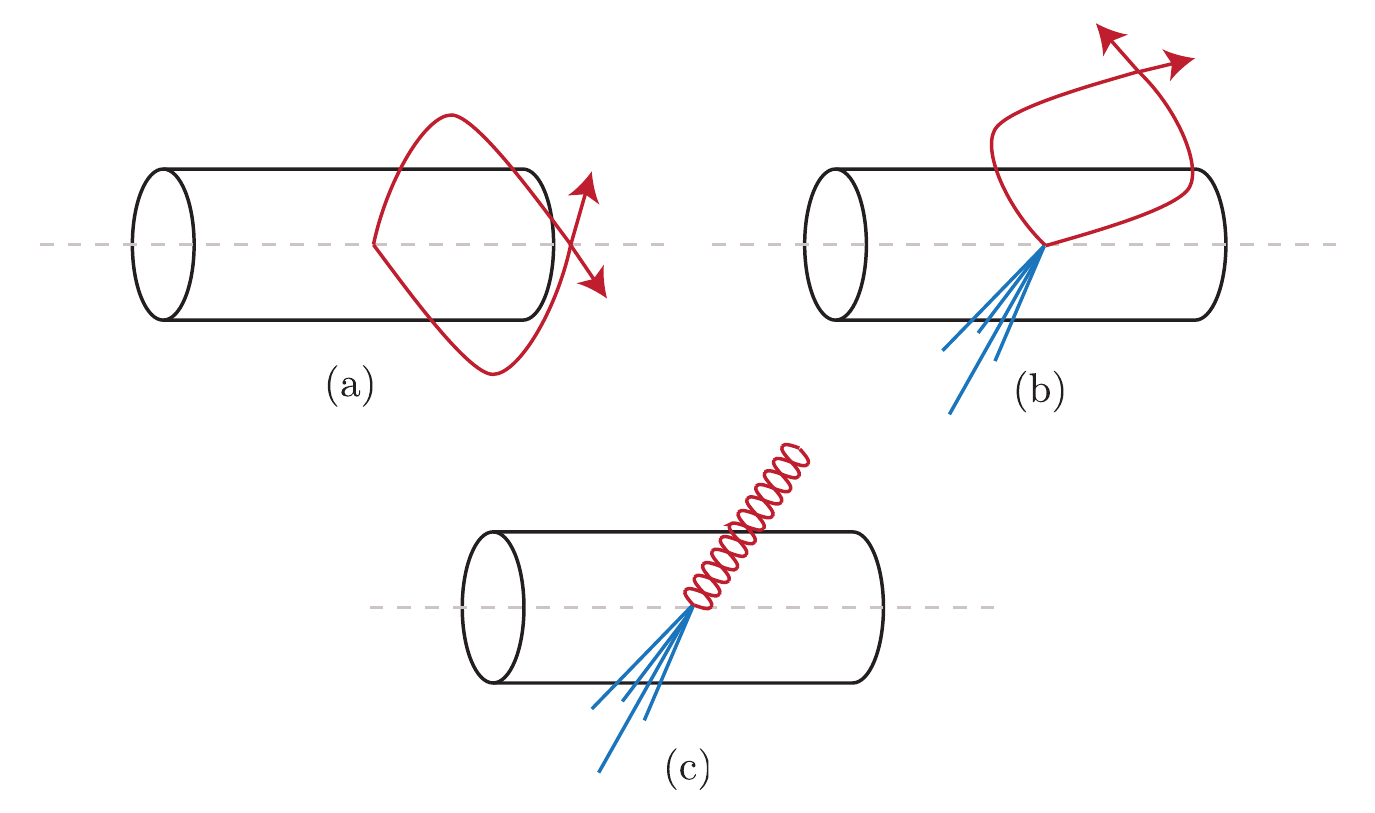}
\caption{Figure 5 from~\cite{Kang:2008ea}. A few examples of the strange paths quirks take through a detector. }
\label{fig:quirk}
\end{centering}
\end{figure}

We can deform the hidden QCD in the following way:
\begin{equation}
\Lambda_{\rm dark} \ll m_Q \sim {\rm TeV}
\end{equation}
where $m_Q$ is the mass of the \textit{lightest} dark quark, and $\Lambda_{\rm dark}$ is the confinement scale of the new gauge group. If these dark quarks also have SM quantum numbers, they can be produced at the LHC, but they then confine with the string length being $\Lambda_{\rm dark}^{-1}$ which can be macroscopic. These objects are termed quirks~\cite{Kang:2008ea}, and they are long lived because it can take a long time for the quirks to find each other to annihilate. They also take strange paths through the detector because the confining string applies a force on them, with a few examples shown in Fig.~\ref{fig:quirk}. The typical signature is then to look for ionizing tracks which do not take the usual path through the detector and instead take some strange oscillating path. Some examples of possible unusual paths are shown in Fig.~\ref{fig:quirk}.

\subsection{Soft Unclustered Energy Patterns (SUEP)}

Another deformation of dark QCD is if the running of the dark gauge coupling, instead of being QCD-like, is approximately conformal and large, then the pattern of energy deposited by the parton shower will be very different. Instead of quarks showering into jet-like structures, the shower will be approximately spherical with a thermal distribution. The majority of hadrons will have very little momentum, with higher energy production simply leading to higher multiplicity. These types of SUEP events, also called soft bombs, were studied in~\cite{Knapen:2016hky}. The most significant challenge for finding these types of models is triggering: with no hard energy, what do you use to discriminate from soft QCD and pile up? Strategies are presented in~\cite{Knapen:2016hky}, and work is ongoing.

\section{Outlook}

Finding the optimum way to design the LHC search program for long-lived particles is still an open question and work is ongoing. There was a workshop in May~\cite{FirstWorkshop} and another planned for October 2017~\cite{SecondWorkshop}. There is a subgroup working in particular on the questions of dark showers, trying to encompass all the models described here. Open questions for theorists include finding ways to simulate all possibilities, including an interpolation between ``emerging jets'' and SUEP.  Open questions for experimentalists include trying to figure out how current jet-finding algorithms in standard searches will react to these sorts of signals.

While it is obvious to many readers, I will stress here that new physics cannot be discovered if it is not triggered on, so developing a triggering strategy with a high efficiency for a wide class of models is critical. The experiments have developed specialized triggers for displaced vertices, but, because the LHC is a hadron collider, triggering on jet activity can sometimes be more efficient than specialized triggers, a point highlighted in~\cite{Csaki:2015fba}.

I will conclude with my personal wishlist for the experimental program on long-lived particles.
\begin{itemize}
\item More searches for distinct collider objects.
\item Searches for different SM states originating in different parts of the detector.
\item More general uses of triggers including multi-jet and VBF. Also, a published list of available triggers and thresholds.
\item Keep searches as model-independent as possible. 
\end{itemize}
Finally I want to thank the organizers of EPS 2017 for inviting me to give this talk and for putting on such an excellent conference.

\end{document}